\documentclass[aps,preprintnumbers,nofootinbib,letter, twocolumn, floatfix]{revtex4} %12pt, showpacs, , twocolumn
\usepackage[english]{babel}
\usepackage{graphicx}% Include figure files
\usepackage{subfig}
\usepackage{hyperref}
\usepackage{dcolumn}% Align table columns on decimal point
\usepackage{slashed}

\newcommand{\newc}{\newcommand}
\newc{\ra}{\rightarrow}
\newc{\lra}{\leftrightarrow}
\newc{\beq}{\begin{equation}}
\newc{\eeq}{\end{equation}}
\newc{\barr}{\begin{eqnarray}}
\newc{\earr}{\end{eqnarray}}

     \newcommand{\GeV}{\,\textrm{GeV}}
\newcommand{\DIR}{./}%{./PLOTS}

\begin{document}

%\rightline{\vbox{\small\hbox{\tt NITS-PHY-2012006} }}
\vskip 1.8 cm

\title{%Odd-sector LHC searches in models with neutrino-like Dark Matter\\
           Limits on electrophilic Dark Matter from LHC Monojets}
\author{K. G. Savvidy}
\affiliation{Nanjing University, Hankou Lu 22, Nanjing, 210098, China \\
College of Science, Nanjing University of Aeronautics and Astronautics, Nanjing 211106, China }
 %%%%%%%%%%%%%%%%%%%%%%%%%%%%%%%%%%%%%%%%%%%%%%%%%%%%%%%%%%%%%%%%%%%%
\begin{abstract} 
%\abstract{
Searches for WIMP Dark Matter particle at the LHC are considered from the point of view of the existence of a Dark Matter particle which couples primarily through the heavy gauge boson Z, as suggested by recent tentative evidence for a 130 GeV gamma line
 in FermiLAT data. 
 %The relevant additional piece is the existence of the second line at 110 GeV, consistent with a monochromatic photon from the Z,H pair annihilation channel. 
 We compare three models in which the WIMP is a neutrino-like particle and consider the limits on such particle and interactions from LHC.

%2) Searches for WIMP Dark Matter particle at the LHC are considered from the point of view of existence of an entire odd sector, which mirrors the quantum numbers of SM particles. This situation takes place in MSSM, in Kaluza-Klein inspired theories with extra dimensions (UED) and in the string-inspired higher-spin model. Each of these models allows for a heavy, neutral partner of the neutrino to be a WIMP, whereas each predict a different spin for such particle, namely MSSM with spin 0 sneutrino, KK UED with spin 1/2 and the higher-spin gauge theory with spin 3/2 neutrino partner. Detailed calculations are provided for all three in the most interesting channels which are dictated by the fact that the odd-sector particle lines are continuous in all Feynman diagrams. We find that the monojet channel, while promising, is perhaps the more difficult in this scenario, while the two lepton channel contribution is elevated above SM.
\end{abstract} 
% \date{\today}
% \pacs{12.60.-i, 95.35.+d, 98.80.Cq}
%%%%%%%%%%%%%%%%%%%%%%%%%%%%%%%%%%%%%%%%%%%%%%%%%%%%%%%%%%%%%%%%%%%%%
\maketitle

\section{Introduction}

Several models of Beyond Standard Model (BSM) physics incorporate a kind of parity which protects the Dark Matter candidate particle (WIMP) of each model from decay \cite{Jungman:1995df,Servant:2002aq}. The same parity demands continuity of Feynman lines of the odd-sector particles in all Feynman diagrams. The lightest odd particle is therefore stable, and if it is to account for Dark Matter must be neutral. One may wish to elevate this idea to a principle to guide construction of physics models, but since models incorporating it already exist we will instead examine how this idea has a unifying power  to guide us in the BSM exotica searches at LHC.

  The natural degree to which this principle is incorporated into the BSM physics models varies. In MSSM, it is the R-parity which is a reasonably natural restriction on possible space of couplings, but violating it does not require giving up either gauge or Lorentz invariance, or renormalizability. In the Kaluza-Klein models, for example UED, the parity is incorporated ad hoc by imposing a boundary condition on wavefunctions in the extra dimension(s) \cite{DNDM}. In the string-inspired higher-spin gauge model \cite{Savvidy:2005fi}, the parity is a natural consequence of requiring all interaction vertices of matter and gauge bosons to be  dimension-4 operators with gauge structure identical to the SM, and Lorentz structure extended with the odd-sector spin 2 gauge bosons and spin 3/2 matter. 
  
  In all these models, the SM is enriched with odd-sector partner particles which have the same gauge quantum numbers as the progenitor SM particles, but have spin lower by 1/2 in SUSY MSSM, equal spin in Kaluza-Klein UED models, and spin higher by integer 1,2,... in the higher-spin model whose origin is stringy. In all models obeying the parity principle, odd-sector particles are created only in pairs, and decay via a chain ending on the WIMP.
  
  If we take the point of view that the Dark Matter WIMP must be an odd partner of some SM particle, it does not yet follow that it is neutrino-like. For example, in MSSM, the LSP is typically a partner of another neutral particle, either a photino \cite{Goldberg} or more generally a neutralino \cite{Kolb:1985nn}. Nevertheless, in this work we suggest LHC searches where the Dark Matter particle is assumed to be a partner of the neutrino and its stability is assured by some parity principle which also implies the existence of an entire odd-sector of partners of the SM particles. Thus, in MSSM there is possibility of a sneutrino WIMP \cite{Falk:1994es, An:2011uq, Belanger:2011ny}. In Kaluza-Klein theory the WIMP may be the heavy partner of the neutrino \cite{Enqvist:1988we, DNDM}, and in  the higher-spin model \cite{Savvidy:2005fi} %,Savvidy:2005zm, Savvidy:2005ki}
it is the spin 3/2 heavy partner of the neutrino. The requirement that the WIMP is neutral strongly limits the possibilities and mass hierarchy in these models \cite{Servant:2002aq, Cheng:2002ab}.

Considerations based on relic abundance disfavor the SM-strength couplings of the neutrino-partner WIMP in MSSM and KK UED model, so that  additional tuning is required \cite{DNDM}. The same considerations in the higher-spin model lead to a range of viable WIMP mass of $110-160 \GeV$ with Majorana WIMP \cite{wVerg}.
 
  We proceed to classify in turn the dominant detection channels at the hadron collider from this point of view.
  
  \section{LHC searches}
   \begin{figure}[!hbp]
\subfloat[Z exchange]{
\includegraphics[height=2.5cm,width=6cm]{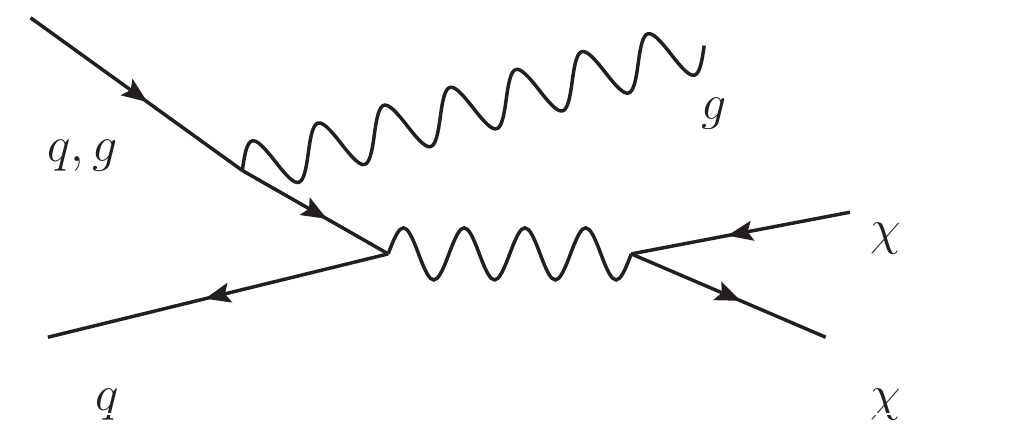} 
} \\
\subfloat[Four-Fermi]{
\includegraphics[height=2.5cm,width=6cm]{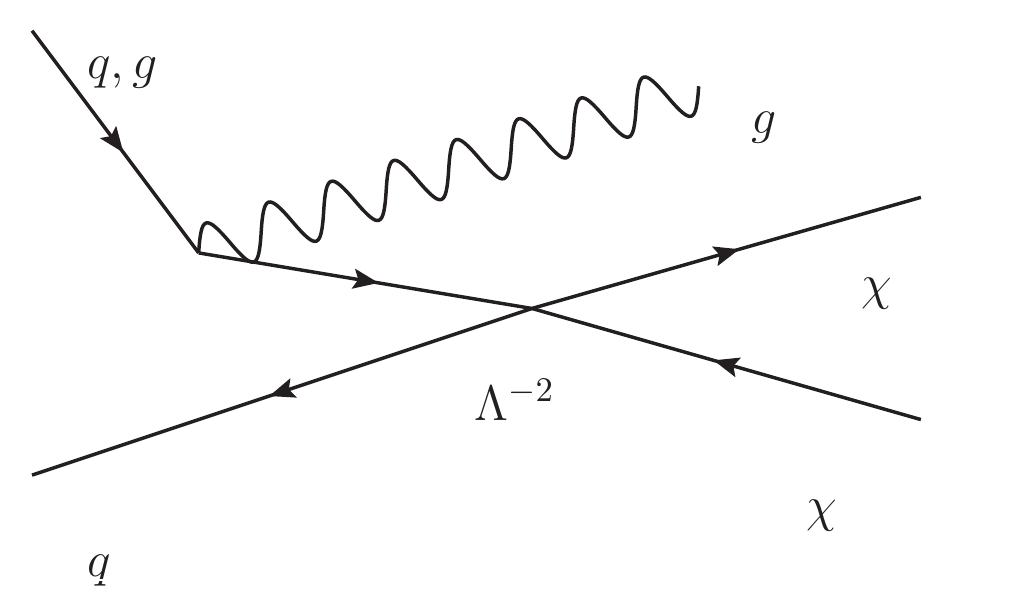}
}
\caption{Tree-level Feynman diagrams for the monojet process. Proceeding via Z in a) and a generic four-fermion effective interaction in b).}
\label{fig:monojetFD}
\end{figure}

{\bf   p,p $\rightarrow$ invisible}
 can proceed via Z, producing a pair of WIMPs which escape undetected. LEP II has placed limit on this up to a mass of $\approx 45 \GeV$ due to the influence of the invisible width on the visible line shape of the Z resonance.  At LHC, and with even much higher assumed WIMP mass, the cross-section is sizable but is observable only in combination with initial state radiation which we consider next.
  
\begin{figure}[h]
\begin{center}
%%\subfloat[SM]     { \includegraphics[height=5cm]{\DIR/plot_SM_Gnn_8TeV_nocut_ang}  }
%\subfloat[SM]     { \includegraphics[height=5cm]{\DIR/plot_SM_Gnn_8TeV_nocut}  }
%%\subfloat[MSSM]{ \includegraphics[height=5cm]{\DIR/plot_MSSM_GXX_msne1000_ang_s}}
%\subfloat[MSSM]{ \includegraphics[height=5cm]{\DIR/plot_MSSM_GXX_mnu130_Pt}}\\
%%\subfloat[KK]     { \includegraphics[height=5cm]{\DIR/plot_KK_GXX_mnu200_ang}  }
%\subfloat[KK-130]     { \includegraphics[height=5cm]{\DIR/plot_KK_GXX_mnu130_Pt}  }
%\subfloat[KK-10]     { \includegraphics[height=5cm]{\DIR/plot_KK_GXX_mnu10_Pt}  }\\
%%\subfloat[HS]    { \includegraphics[height=5cm]{\DIR/plot_HS_GXX_mnu130_ang} }
%\subfloat[HS]    { \includegraphics[height=5cm]{\DIR/plot_HS_GXX_mnu130_Pt} }
\includegraphics[height=5cm]{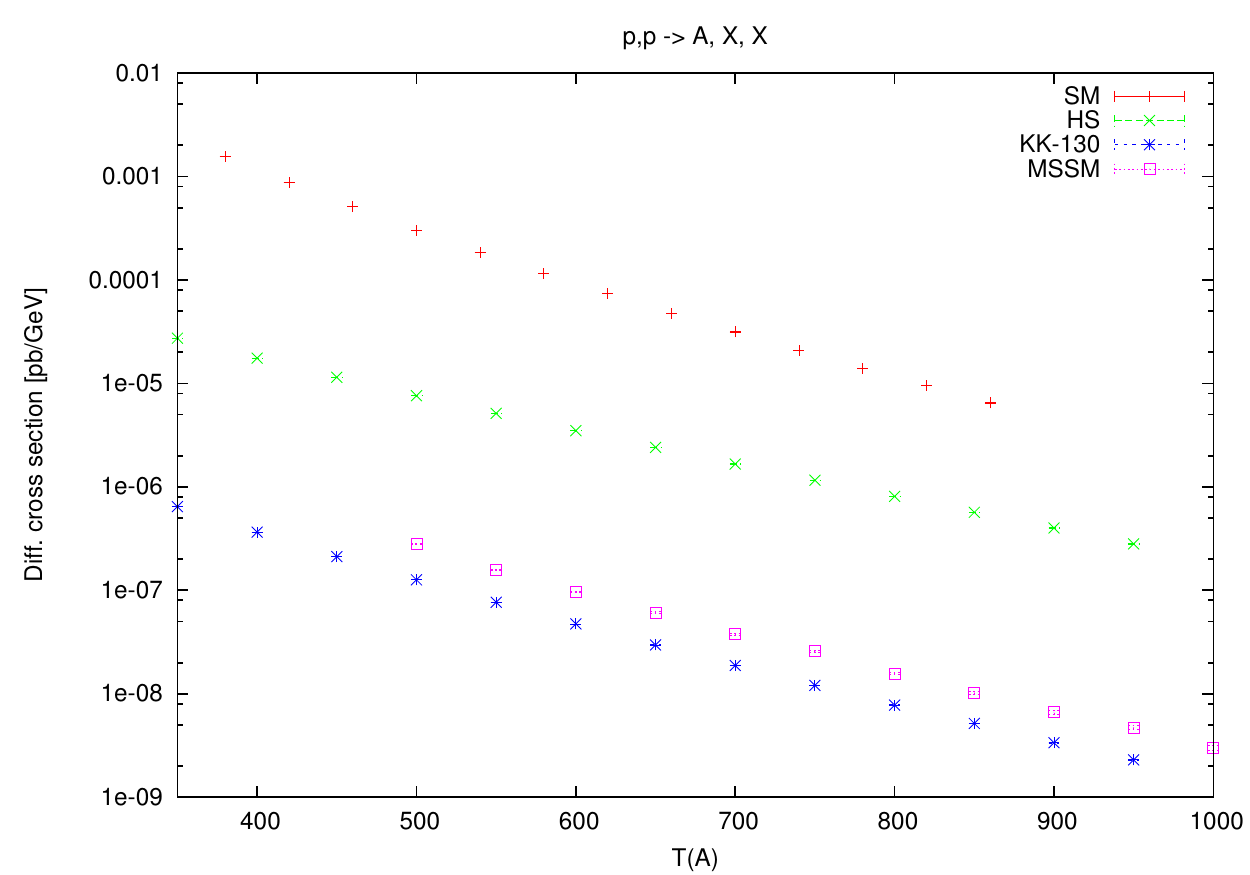}
\caption{The contributions to the monojet final state. ТTransverse momentum distributions for %on the right.
a) SM b) SUSY c) KK UED d) higher spin model. Cuts on $Pt > 350\GeV$ and $|\eta|<2.4$}
\label{fig:monojet}
\end{center}
\end{figure}

{\bf  p,p $\rightarrow$ 1 jet or photon + $\slashed{E}_T$}
 has emerged as the golden channel for DM search at LHC \cite{fox}. Feynman diagram typically responsible for this is Fig. \ref{fig:monojet} a) and b).
%\clearpage
This channel is currently analyzed in a model-independent way \cite{max,tait,fox},  in terms of an effective four-fermion interaction as in Fig. \ref{fig:monojet} b). The experiment places a combined limit on the WIMP mass and the strength of the effective interaction $\Lambda$. This interaction presumably requires existence of a new boson of mass $M_X \approx 1/\sqrt{\Lambda}$ to carry this interaction, the current limits on $M_X$ and $M_\chi$ being excluded in a more or less rectangular region  up to $\approx 800 \GeV$ by recent ATLAS data \cite{ATLASmonojet}.

From the parity-principle point of view, the interaction cannot be due to an odd-sector boson (at least at tree-level), so that Z exchange is the preferred possibility. Unfortunately, in this regime, Z acts like a light particle so that any effective four-fermion description is not possible. The WIMP pair creation channel gives only a small excess over the dominant and irreducible SM background which is $p,p \rightarrow j,\nu,\bar{\nu}$. In this case, the experimental limits are weakened dramatically, as already argued by the authors of \cite{fox}. It is not clear that the systematic uncertainties will allow for absolute calibration of this channel to 3-5\% accuracy which is required to make a discovery, no matter what the ultimate integrated luminocity will be achieved. Total acceptance-corrected cross sections are presented in Table \ref{tab:xs} at 8 TeV and Table \ref{tab:xs1000} at 13 TeV. Specific predictions for the $\slashed{E}_T$ distribution of events is given in Fig \ref{fig:monojet} in the UED spin 1/2 and higher spin model for spin 3/2 neutrino-like WIMP.

\section{Cosmological Abundance}
In MSSM toy model with only a slepton partner multiplet with all interactions dictated by the gauge principle, or in the full
 MSSM, in the corner of parameter space where only sneutrino and selectron play a role, the abundance looks like this:
\begin{figure}[h]
\begin{center}
\subfloat[MSSM, sneutrino with spin 0]{ \includegraphics[height=6cm]{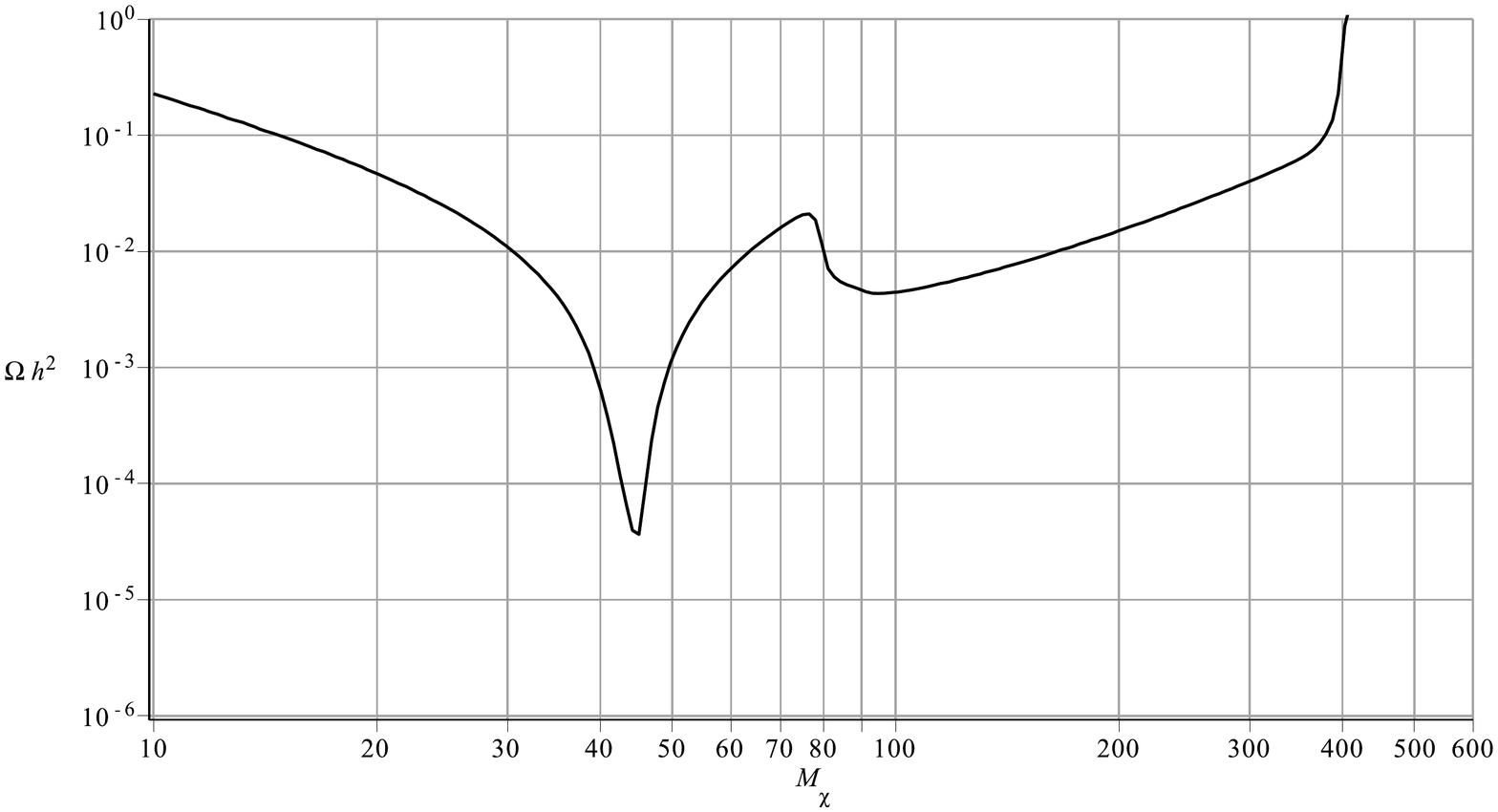}}\\ 
%complex sneutrino with spin 0
\subfloat[KK, majorana neutrino with spin 1/2]    { \includegraphics[height=4cm]{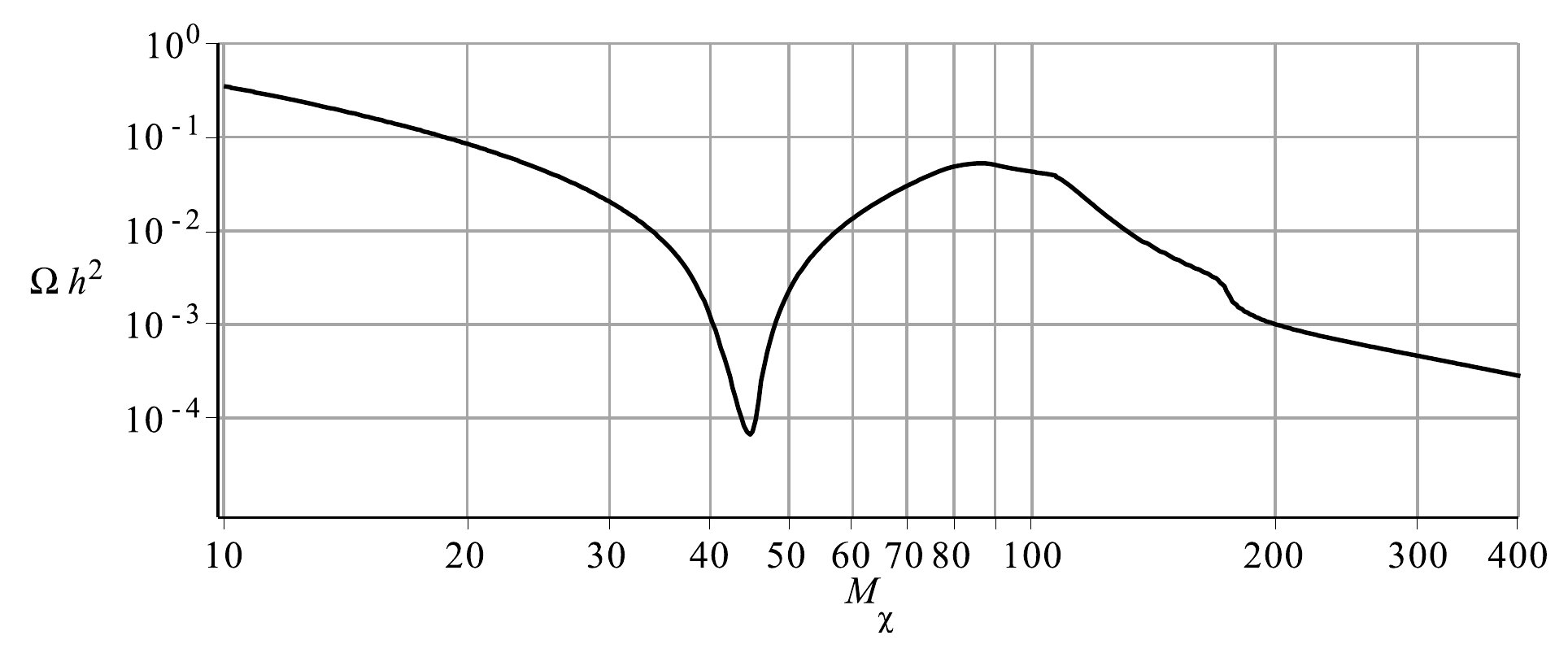}  }\\
\subfloat[HS, majorana neutrino with spin 3/2]    { \includegraphics[height=4cm]{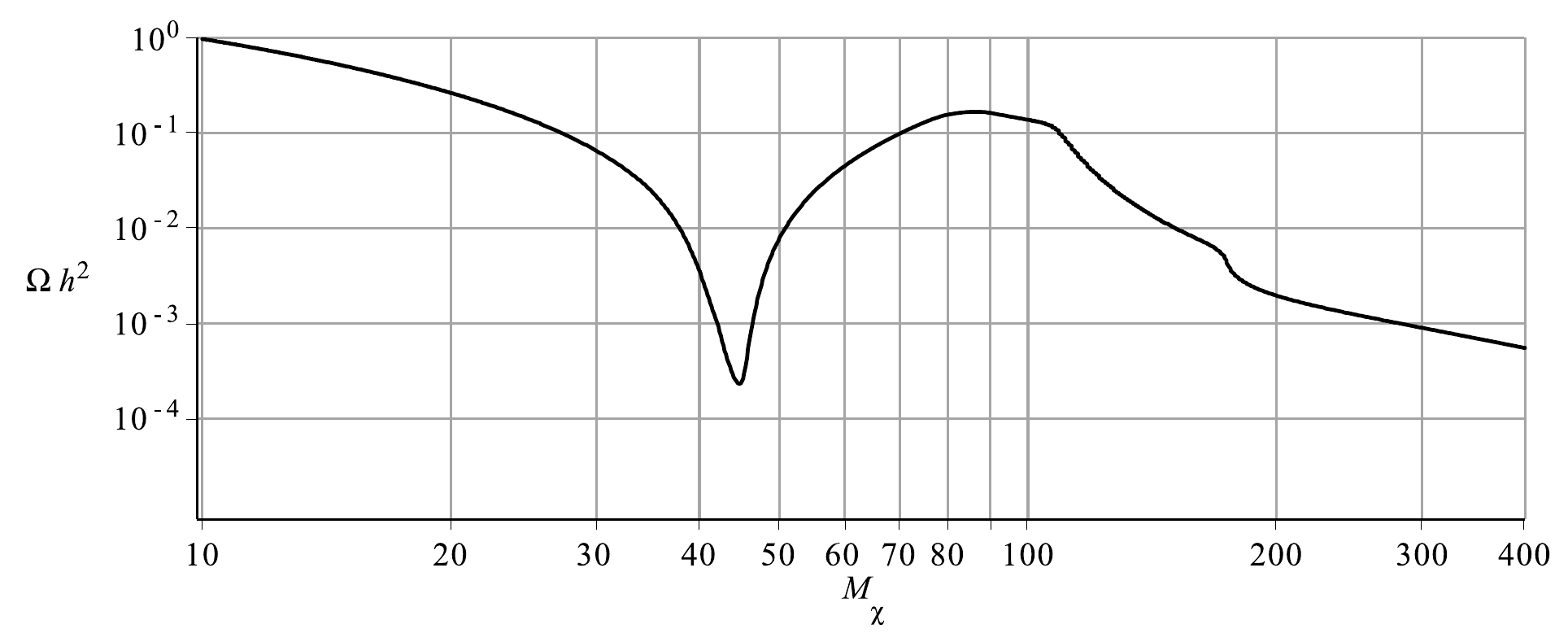} }
\caption{Cosmological abundances calculated in 
a) MSSM b) KK d) higher spin model (HS)}
\label{fig:abundance}
\end{center}
\end{figure}

Recently, some models of sneutrino Dark Matter have emerged in which a left-right symmetric extension is assumed, such that the sneutrino is a viable Dark Matter candidate if it is predominantly consists of the right-handed, sterile component \cite{An:2011uq}. In our analysis, it is assumed that the lightest s-neutrino is predominantly of a single flavor and that co-annihilation with other MSSM particles does not play a role. It is apparent from from the plot that the abundance is sizable $\Omega h^2 > 10^{-2}$ for a range of mass roughly in the region 60-400 GeV, and in this region the toy model can be extended with the right-handed sterile s-neutrino such that the mixing provides a lower coupling to the Z, due to a mixing angle which is not unnaturally small. Nevertheless, we do not in fact make this elaboration,  so that the left-handed neutrino couples to the Z with full strength, which makes our collider result strictly an upper bound.

In the case of the KK Dirac heavy spin 1/2 neutrinos, the abundance does not reach the cosmologically relevant value \cite{Enqvist:1988we}, but in the Majorana case it is typically larger \cite{Goldberg, Kolb:1985nn, Griest:1988ma} and comes close near 80-90 GeV as in Fig. \ref{fig:abundance}b.

Furthermore, it was found in \cite{wVerg} that the abundance in the case of Majorana spin 3/2 neutrino-like WIMP reaches the cosmological value and is a viable candidate in the range of mass 70-160 GeV which allows for the recently discovered 130 GeV monochromatic line by Fermi-LAT. The direct detection limits in this case are satisfied, and it was found that the current generation of Xenon experiments are sensitive enough to confirm or rule out this model due to their content of odd-neutron isotopes.

\section{Results}
\label{sec:results}
The Monte-Carlo generation of events was done with calchep, cteq6m distributions and at 4+4=8TeV.\\
{\bf monojet}
The total crosssection of the irreducible SM contribution with $Pt > 350\GeV$ and $|\eta|<2$  is 
$\sigma \times A (p,p \to j, \nu,\bar{\nu}) = 231$ fb.

This crosssection is reported after including the cuts so acceptance is taken into account, but not efficiency, 
 which is in any case high at 83\% average as reported by ATLAS.

\begin{table}[h]
\caption{The various mono-jet crosssections with  cuts on $Pt > 350\GeV$ and $|\eta|<2.0$ corresponding to the ATLAS Signal Region SR3 at 8TeV. All of these numbers should be scaled down by efficiency which is reported by ATLAS to be around 83\% in their detector and parameters of their analysis. The estimate on the systematical error is based on extrapolation of the 2011 ATLAS analysis.} 
\label{tab:xsJ}
\begin{center}
\begin{tabular}{||l|c|c|c|}
\hline
Model & $M_\chi$ & $\sigma \times$ A &  Events with 20 fb$^{-1}$\\
\hline
SM      & 0     & 231 fb                &                $4600 \pm 70$ (stat) $\pm 314$ (syst)\\ 
KK-Dirac      & 10   & 18.8 fb     &                380\\
KK-Dirac      & 130 & 0.077 fb   &                $1$\\
KK-Dirac      & 200 & 0.027 fb   &                $<1$\\
KK-Majorana& 130 & 0.012 fb   &                $<1$\\
MSSM          & 104 & 0.16 fb    &                $3$\\
HS                & 130 & 4 fb          &                80\\ 
\hline
\end{tabular}
\end{center}
\end{table}
%KK monojet with  $Pt > 350\GeV$, $|\eta|<2.4$, $M_\chi=200$GeV is $\sigma \times A (p,p \to g, \chi,\bar{\chi}) =0.027$ fb.
%KK monojet with  $Pt > 350\GeV$, $|\eta|<2.4$, $M_\chi=130$GeV is $\sigma \times A (p,p \to g, \chi,\bar{\chi}) =0.077$ fb.
%HS monojet with $Pt > 350\GeV$, $|\eta|<2.4$, $M_\chi=130$GeV is $\sigma \times A (p,p \to g, \chi,\bar{\chi}) =4$ fb.

The immediate conclusion from this table is that the data can be used to exclude an extra light neutrino-like particle at least in some  range of mass, but barely competitive with the LEP II limit of 45 GeV.
\begin{figure}[!h]
\begin{center}
\includegraphics[height=5cm]{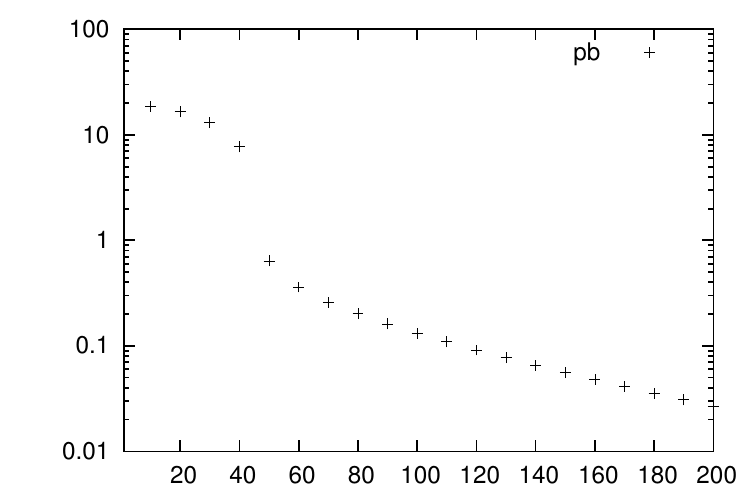}
\caption{Crosssection as a function of the KK neutrino mass. All with the same cuts as in Fig \ref{fig:monojet} and Table \ref{tab:xsJ}.}
\label{fig:KKdependence}
\end{center}
\end{figure}

In the higher-spin gauge model, the massive spin 3/2 particle can have a harder tail at high-Pt due to its Lorentz boost properties, to some extent mimicking the behavior of the cross section in the effective four-fermion theory.

\begin{figure}[htbp]
\begin{center}
%%\subfloat[SM]     { \includegraphics[height=5cm]{\DIR/plot_SM_gammaNN_Pt150_ang}  }
%\subfloat[SM/1.5]     { \includegraphics[height=5cm]{\DIR/plot_SM_gammaNN_Pt150}  }
%%\subfloat[MSSM]{ \includegraphics[height=5cm]{\DIR/plot_MSSM_gammaNN_msne300_ang}}
%\subfloat[MSSM]{ \includegraphics[height=5cm]{\DIR/plot_MSSM_gammaNN_msne104_Pt}}\\
%%\subfloat[KK]     { \includegraphics[height=5cm]{\DIR/plot_KK_gammaNN_mnu300_ang}  }
%\subfloat[KK]     { \includegraphics[height=5cm]{\DIR/plot_KK_gammaNN_mnu130_Pt}  }
%%\subfloat[HS]    { \includegraphics[height=5cm]{\DIR/plot_HS_gammaNN_Mn130_ang} }
%\subfloat[HS]    { \includegraphics[height=5cm]{\DIR/plot_HS_gammaNN_Mn130} }\\
\includegraphics[height=5cm]{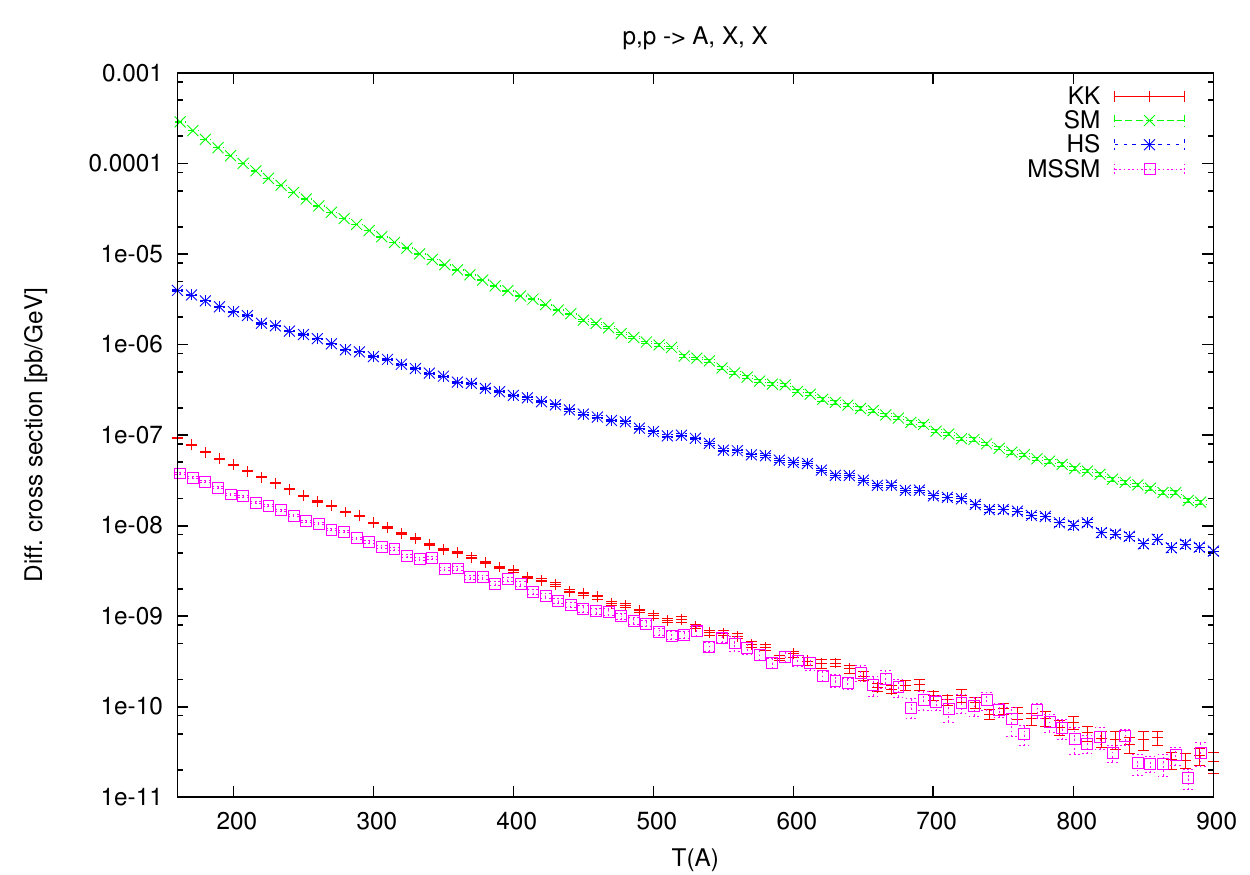}
\caption{The contributions to the monophoton final state. 
 SM - green, SUSY - pink,  KK  UED - red,  higher spin model - blue.}
\label{fig:monophoton}
\end{center}
\end{figure}

\begin{table}[h]
\caption{The various mono-photon cross sections with  cuts on $Pt > 150\GeV$ and $|\eta|<2.4$.} 
\label{tab:xsP}
\begin{center}
\begin{tabular}{||l|c|c|c|}
\hline
Model & $M_\chi$ & x-section \\
\hline
SM & 0     & 28.8 fb  \\ 
HS & 130 &  0.4 fb    \\ 
KK & 130 & 0.0075 fb \\
MSSM&104&0.016 fb \\
\hline
\end{tabular}
\end{center}
\end{table}
%The total crosssection of the SM contribution with $Pt > 150$ and $|\eta|<2.4$ (as in CMS) is 
%$\sigma \times A (p,p \to \gamma, \nu,\bar{\nu}) = 19.25$ fb . 
%
%The total crosssection of the spin 3/2 contribution with $Pt > 150$ and $|\eta|<2.4$ (as in CMS) is 
%$\sigma \times A (p,p \to \gamma, \nu,\bar{\nu}) = 0.4$ fb.

\section{LHC at 13 TeV}
    Here we examine the likely situation at the upgraded LHC operating at the planned center of mass energy of 13 TeV. It is expected that the ultimate integrated luminosity may exceed 100 fb$^{-1}$. We first examine the predictions in this regime for the irreducible Standard Model process $Z\to \nu, \bar{\nu}$ with all three neutrino flavors. The cross section times acceptance at this energy and with a Pt cut of 500 GeV is similar that at 8 TeV and Pt cut of 350 GeV.
 
\begin{table}[h]
\caption{The various mono-jet cross sections with  cuts on $Pt > 500\GeV$ and $|\eta|<2.0$. Included is the projected statistical and systematical uncertainty. The systematical uncertainty is extrapolated using the 2011 results and assuming reduction of systematical uncertainty which is due to limitations of MC modeling.} 
\label{tab:xs}
\begin{center}
\begin{tabular}{||l|c|c|c|}
\hline
Model & $M_\chi$ & x-section & Events with 100 fb$^{-1}$\\
\hline
SM & 0     & 159.5fb     & $16000 \pm 130$ (stat) $\pm 1100$ (syst)\\ 
HS & 130 & 15.8 fb       & $1600 \pm 40$ (stat)\\ 
KK & 130 & 0.097 fb & $10 \pm 3$ (stat) \\
MSSM& 104 & 0.2 fb & $200 \pm 15$ (stat)\\
\hline
\end{tabular}
\end{center}
\end{table}
%SM monojet with  $Pt > 500\GeV$, $|\eta|<2.0$,  $\sigma \times A (p,p \to g, \chi,\bar{\chi}) =159.5$ fb.
%KK monojet with  $Pt > 500\GeV$, $|\eta|<2.0$, $M_\chi=130$GeV is $\sigma \times A (p,p \to g, \chi,\bar{\chi}) =0.097$ fb.
%HS monojet with $Pt > 500\GeV$, $|\eta|<2.0$, $M_\chi=130$GeV is $\sigma \times A (p,p \to g, \chi,\bar{\chi}) =15.8$ fb.
This cut corresponds to the ATLAS Signal Region SR4 in the analysis of the 2011 LHC run with 5fb$^{-1}$.

\begin{figure}[h]
\begin{center}
%\subfloat[SM]     { \includegraphics[height=5cm]{\DIR/plot_SM_Gnn_8TeV_nocut_ang}  }
%\subfloat[SM]     { \includegraphics[height=5cm]{\DIR/plot_SM_Gnn_13TeV_Pt500}  }
%\subfloat[KK]     { \includegraphics[height=5cm]{\DIR/plot_KK_GXX_13TeV_mnu130_Pt}  }\\
%\subfloat[HS]    { \includegraphics[height=5cm]{\DIR/plot_HS_GXX_13TeV_mnu130_Pt} }
\includegraphics[height=5cm]{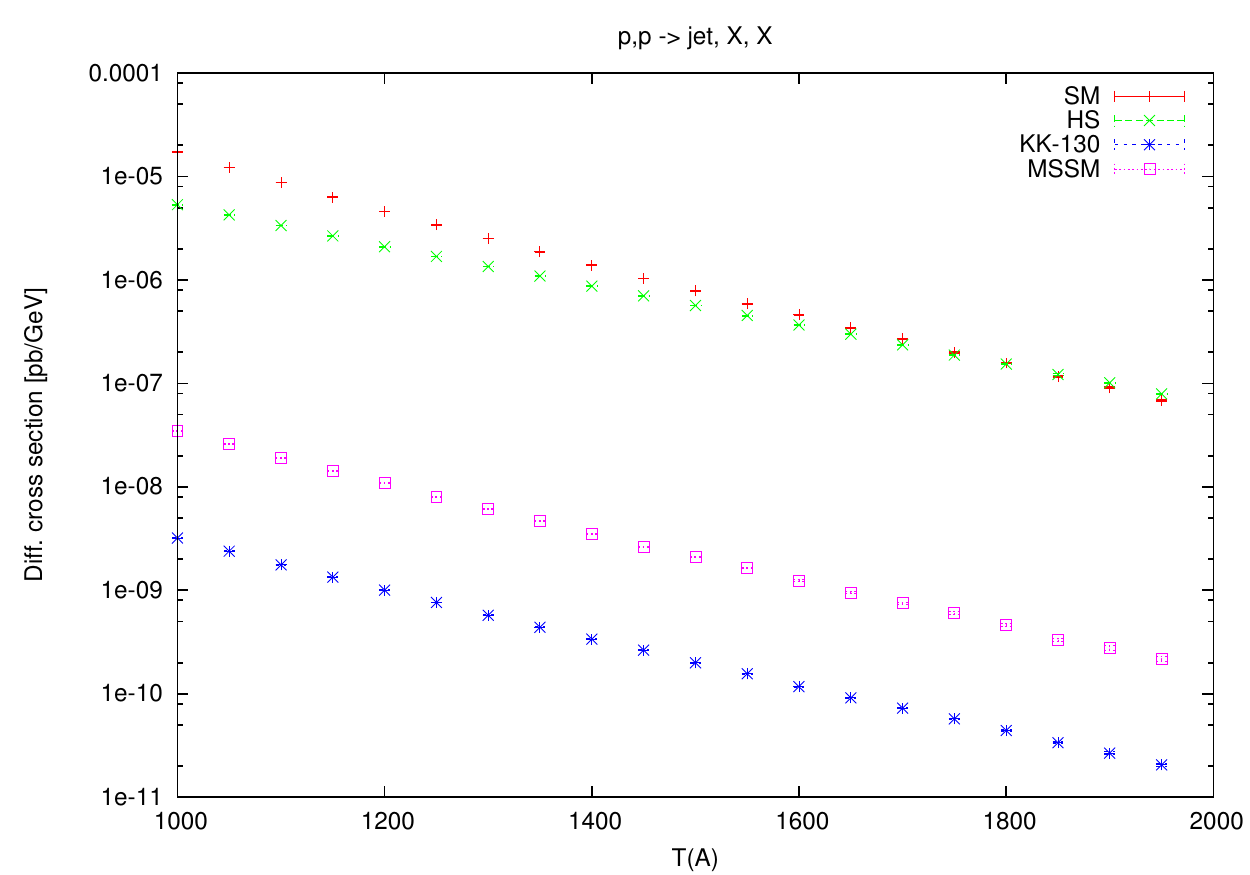}
\caption{The contributions to the monojet final state at 13 TeV. Transverse momentum distributions.
%a) SM b) SUSY c) KK UED d) higher spin model
}
\label{fig:monojet13TeV}
\end{center}
\end{figure}

Ultimate reach with 100fb$^{-1}$ is perhaps with $Pt > 1000\GeV$ where the cross sections times acceptance are in Table \ref{tab:xs1000}.
\begin{table}[h]
\caption{The various mono-jet crosssections at 13 TeV with  cuts on $Pt > 1000\GeV$ and $|\eta|<2.0$.} 
\label{tab:xs1000}
\begin{center}
\begin{tabular}{||l|c|c|c|}
\hline
Model & $M_\chi$ & x-section & Events with 100 fb$^{-1}$\\
\hline
SM & 0     & 3.2fb    (3spec)  & $320 \pm18$ (stat) $\pm 35$ (syst)\\ 
HS & 130 & 1.3 fb       & $130 \pm12$ (stat)\\ 
KK & 130 & 0.0035 fb & $<1$\\
MSSM& 104 & 0.007 fb &$<1$\\
\hline
\end{tabular}
\end{center}
\end{table}
%SM monojet with  $Pt > 1000\GeV$, $|\eta|<2.0$,  $\sigma \times A (p,p \to g, \chi,\bar{\chi}) =3.2$ fb.
%KK monojet with  $Pt > 1000\GeV$, $|\eta|<2.0$, $M_\chi=130$GeV is $\sigma \times A (p,p \to g, \chi,\bar{\chi}) =1.3$ fb.
%HS monojet with $Pt > 1000\GeV$, $|\eta|<2.0$, $M_\chi=130$GeV is $\sigma \times A (p,p \to g, \chi,\bar{\chi}) =0.0035$ fb.

\section{Conclusions}

It seems that the case of a heavy WIMP with coupling to matter primarily through the Z is not very favorable for discovery in  the monojet and monophoton channel, since there is now evidence that the precision of a hadron collider is not sufficient for this no matter how large is the total integrated luminosity. The optimistic situation arises only if the WIMP has for one reason or another a hard tail in Pt. One is the four-fermi effective interaction which has been extensively studied, and the other, considered in this paper, where the WIMP is a spin 3/2 partner of the neutrino.
% Second, that the two lepton plus missing energy channel should be analysed in terms of {\em pair} creation of $W^\prime$s, both decaying into a lepton and a WIMP. Here the situation is more optimistic, especially if this branching ratio is 1 leading to elevation of this interesting signal above the irreducible SM background.

\vfill

\end{document}